\documentclass[conference]{IEEEtran}
\IEEEoverridecommandlockouts

\usepackage[ruled,vlined,linesnumbered]{algorithm2e}
\usepackage{algpseudocode}
\usepackage{amsmath}
\usepackage{amssymb}
\usepackage{amsfonts}
\usepackage{authblk}
\usepackage{balance}
\usepackage{booktabs} 
\usepackage{cite}
\usepackage{etoolbox}
\usepackage{enumitem}
\usepackage{graphicx}
\usepackage{verbatim}
\usepackage{multirow}
\usepackage{xcolor}
\usepackage{xspace}

\def\BibTeX{{\rm B\kern-.05em{\sc i\kern-.025em b}\kern-.08em
    T\kern-.1667em\lower.7ex\hbox{E}\kern-.125emX}}
    
\newcommand{\ie}{i.e.\xspace}
\newcommand{\etal}{\textit{et. al}\xspace}
\newcommand{\eg}{e.g.\xspace}

\newtheorem{mydef}{\textbf{Definition}}

\newcommand{\tool}{\textsc{$A^{3}$Ident}\xspace}
\begin{document}

\title{\tool: A Two-phased Approach to Identify the Leading Authors of Android Apps}

\author[1*]{Wei Wang\thanks{*~the authors contributed equally to this work}}
\author[2,3*]{Guozhu Meng}
\author[4]{Haoyu Wang}
\author[2,3]{Kai Chen}
\author[1]{Weimin Ge}
\author[1]{Xiaohong Li}

\affil[1]{Tianjin Key Laboratory of Advanced Networking (TANK), School of Computer Science and Technology,  College of \authorcr Intelligence and Computing, Tianjin University, Tianjin, China, \{wei2018, gewm, xiaohongli\}@tju.edu.cn}
\affil[2]{Institute of Information Engineering, Chinese Academy of Sciences, China, \{mengguozhu, chenkai\}@iie.ac.cn}
\affil[3]{School of Cyber Security, University of Chinese Academy of Sciences, China}
\affil[4]{Beijing University of Posts and Telecommunications, Beijing, China, haoyuwang@bupt.edu.cn}

\maketitle
\thispagestyle{plain}
\begin{abstract}
Authorship identification is the process of identifying and classifying authors through given codes. Authorship identification can be used in a wide range of software domains, \eg, code authorship disputes, plagiarism detection, exposure of attackers' identity. 
Besides the inherent challenges from legacy software development, framework programming and crowdsourcing mode in Android raise the difficulties of authorship identification significantly. More specifically, widespread third party libraries and inherited components (\eg, classes, methods, and variables) dilute the primary code within the entire Android app and blur the boundaries of code written by different authors. 
However, prior research has not well addressed these challenges.

To this end, we design a two-phased approach to attribute the primary code of an Android app to the specific developer. In the first phase, we put forward three types of strategies to identify the relationships between Java packages in an app, which consist of context, semantic and structural relationships. 
A package aggregation algorithm is developed to cluster all packages that are of high probability written by the same authors. 
In the second phase, we develop three types of features to capture authors' coding habits and code stylometry. 
Based on that, we generate fingerprints for an author from its developed Android apps and employ several machine learning algorithms for authorship classification. 
We evaluate our approach in three datasets that contain 15,666 apps from 257 distinct developers and achieve a 92.5\% accuracy rate on average. 
Additionally, we test it on 2,900 obfuscated apps and our approach can classify apps with an accuracy rate of 80.4\%.

\end{abstract}

%
%

\section{Introduction}\label{sec:intro}

Code authorship identification is a generic technique of determining the author for a specific piece of code. It has been widely used in multiple areas including code authorship dispute, plagiarism detection~\cite{plagiarismdetection}, app clone detection~\cite{appauth}, software forensics~\cite{softwareforensics}, and malware analysis~\cite{malwareanalysis,issta2016smart}. 
Taking malware analysis~\cite{malwareanalysis2} in Android as an example, the cost of making and evolving malware is relatively low due to automated code generation techniques~\cite{appgens} and amounts of reusable code. As a consequence, manually analyzing these malware samples becomes a laborious and tedious task when anti-malware tools~\cite{antimalware,antimalware2} cannot effectively capture malicious behaviors inside. By applying authorship identification, security analysts can determine the author of malware and further infer the contained malicious behaviors and attack targets.

Software developers usually leave personal and identifiable information on the code in terms of distinguishable programming habits. This information is the pivot for accurately identifying its author. However, it is very challenging due to the scarcity of datasets, evolving programming style, code obfuscation, etc~\cite{authorship2019csur}.
The research on code authorship of modern software can date back to the 1980s, where Oman and Cook employed typographic characteristics to distinguish Pascal programs~\cite{Oman}. 
Subsequently, more stylistic features have been extracted and utilized for authorship identification. 
Lexical and syntactic features of source code, \eg, variable naming, layout style, and the use of data structure can de-anonymize the authors of code ~\cite{AbuhamadAMN18,Frantzeskou,ngrams}. Semantic features such as an abstract syntax tree, control flow graph, and program dependence graph~\cite{bin2016fse,SimkoZK18 } are also found effective in authorship identification. 
However, these techniques cannot be fully applied in attributing authors of Android apps owing to the distinct development mode.

Android app is a combination of functional modules, and its code is written by following the development rules by either an individual or a team. Therefore, the first challenge comes from the code influence of the collaborative group. To be more specific, Android apps may be developed in teams, and the different modules can be designed by multiple authors. All the legacies from the teamwork dilute authors' programming styles. The second challenge is due to numerous reusable libraries for easing development difficulty~\cite{libscout,libd,commonlibraries}. 
For example, developers are prone to use advertisement libraries (\eg, \emph{AdMob} and \emph{Facebook}), social networks (\eg, \emph{Twitter} and \emph{Wechat}), development libraries like \emph{okHttp} and \emph{Google GSON}. According to an empirical study on 100 F-Droid apps, we found that only 8\% of them were produced independently by developers without any third-party libraries. 
As a consequence, the introduction of third-party libraries certainly exerts a negative influence on authorship identification. 
Last, Android apps do not fully retain lexical and syntactic features of source code after compilation, constituting the third challenge. Besides, the use of \textsc{ProGuard} during compilation can obfuscate source code drastically. The studies \cite{Kalgutkarauthorship,authorship2018codaspy} make the pioneering efforts to distill recognizable features like strings, used data structures, and statistics of methods and classes from an apk file. However, all of them fail to address all the three challenges as aforementioned. Hence, we aim at studying segmented pieces of the app code instead of the entire code to solve the challenge.

In this study, we propose a two-phased approach, termed as \tool, to identify the leading authors for Android apps which consists of \emph{authorship decoupling} and \emph{authorship identification}. The leading author is an Android developer who primarily implements the advertised functional model, \ie, the primary module~\cite{PiggyApp}.
In the first phase, we construct a package relation graph for a given app, and group all packages that are likely created by one single author based on three assumptions (see Section~\ref{sec:apkfilter}). 
We propose semantic and structural similarities to quantify the distance between packages of the app. The packages are further aggregated with the \emph{Louvain} model~\cite{louvain}. In this manner, we can identify the primary module created by the leading author and address the one and second challenges. 
In the second phase, we extract three types of styling features and get rid of problems brought by the Android framework and obfuscation, such as overloaded methods, identifier rules, etc (see Section~\ref{sec:authorshipIdentification}). For these extracted features, we employ \emph{word2vec}~\cite{word2vec} to embed authors' profiles. At last, we use three machine learning models (\ie, Linear SVM~\cite{linearsvm}, Random Forest~\cite{randomforest}, and Logistic Regression~\cite{regression}) to determine the possible authors for the test apps. \tool is evaluated extensively on four datasets: F-Droid, benignware, malware, and obfuscated apps. 
As indicated by experimental results, \tool has achieved an accuracy of 96\%, 98.9\%, 82.7\%, and 80.4\%, respectively. By comparing an open-source authorship attribution tool \textsc{AppAuth}~\cite{appauth}, our approach makes a 3.4\% improvement in authorship identification with an accuracy of 87.8\%. 
In addition, we identify four types of external code that are mixed with the primary code during app development and compilation. 
The experiment on obfuscated apps proves that our approach stays a high accuracy in authorship identification against obfuscation. 

\noindent\textbf{Contributions.} 
We have made the following contributions.
\begin{itemize}[leftmargin=*]
	\item We propose authorship decoupling in the granularity of package to group the correlated code as per its authors, which is never considered in prior research on Android. Based on this, we find that several classes have been integrated into the apk file during compilation, \eg, configuration classes and third-party libraries. 
	\item For authorship identification, we extract three types of features from the primary module, \ie, dex-level, lib-level, and manifest-level features. Differently, we eliminate the influence of the Android framework of feature extraction and combine \emph{TF-IDF} and word2vec techniques to embed the features. Three classifiers are then employed for identifying authorship. 
	\item We implement an automated tool \tool which is extensively evaluated on 257 authors with their 15,666 apps. Authorship decoupling achieves an accuracy of 96.11\% on 416 F-Droid apps, and authorship identification achieves a 92.5\% accurate rate on average on the whole set. Our approach proves to be also effective in handling obfuscated apps with only a 7.2\% reduction in accuracy.
\end{itemize}
\section{Background}\label{sec:background}

\subsection{Android app Authorship} 
Android developers compile their source code and other resource files, \eg, layout files, into an Android application package (APK) and deliver it into the Android platform. 
An apk file contains \textsf{\small AndroidManifest.xml}, \textsf{\small *.dex}, \textsf{\small *.arsc}, \textsf{\small res} directory, \textsf{\small META-INF} directory, etc. Developers are required to sign Android apps with their own certificates. 
The signing certificate is stored in the ``\textsf{\small META-INF}'' folder. 
Android signatures guarantee the integrity of apps and prevent tampering and replacement. Except for certificates are exposed in the Android source code published by AOSP~\cite{publicKey} and the private key is somehow leaked~\cite{privatekey}, developers' certificates are confidential and cannot be known by others~\cite{sign2019ase}. Hence, apps signed with the same certificate are supposed to be created by the same developer.

The study of authorship identification stems from the field of literature, which aims to identify the author of a controversial text based on his/her unique linguistic styles (\eg, verb, vocabulary, sentence length). It is also very significant in the domain of Android. For example, authorship identification can assist the confirmation of adversaries' identity behind zero-day attacks or variants in the wild. Generally, it elicits a set of characteristics as fingerprints to quantify the stylometry (\eg, programming styles, and naming conventions) of malware authors. 
Different from other software systems, one apk file is most likely composed of multiple pieces of code from different authors and compilation. As a result, we have to separate the primary code from external code.

\subsection{Louvain model} 
Louvain model is a graph clustering algorithm based on multi-level modularity optimization to identify communities from a large network, which we use for authorship decoupling (see Section~\ref{relatcorr}). 
The nodes and the weight between nodes as input, Louvain model outputs the cluster to which each node belongs.  
Given a weighted graph, its goal is to maximize the modularity guided by $Q$ as follows~\cite{louvain}:
\begin{flalign}&
Q=\frac{1}{2m}\ast \sum_{ij}\left [ A_{i,j}-\frac{k_i\ast k_j}{2m} \right ]\ast \sigma(C_i,C_j) \label{eqt:louvain} 
\end{flalign}
where $Q$ denotes the current status of modularity, \(A_{ij}\) represents the weight between node i and node j, \(k_i\) represents the weighted sum of all edges connected with node i, \(C_i\)  is the cluster number of node i, and \(\sigma(C_i,C_j)\) represents that if node i and j are in the same cluster, the value is 1, otherwise the value is 0. At first, each node is treated as an independent category, and Louvain algorithm is divided into two steps:
\begin{enumerate}[leftmargin=*,label=Step \arabic*]
	\item Traversing all nodes of the graph, grouping similar nodes together, and labeling them until the cluster of all nodes does not change.
	\item Re-initializing the graph and merging the nodes from the same cluster into a supernode. The supernode contains a self-connected edge whose weight doubles the sum of weights of all edges in the original. The weight between supernodes is the sum of weights of the edges whose connected nodes locate different clusters. Then, repeat the first step until the modularity $Q$ does not change. 
\end{enumerate} 
\section{Authorship Analysis and System Overview}\label{sec:approach}

In this section, we conduct an empirical analysis of Android apps and then present the system overview of \tool.
\begin{figure}
	\centering
	\includegraphics[width=0.35\textwidth]{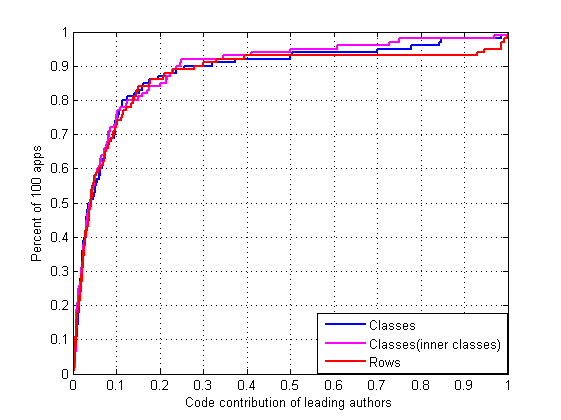}
	\caption{The distribution of code contribution of authors.}
	\label{fig:codepercent}
\end{figure}

\begin{figure*}
	\centering
	\includegraphics[width=.8\textwidth]{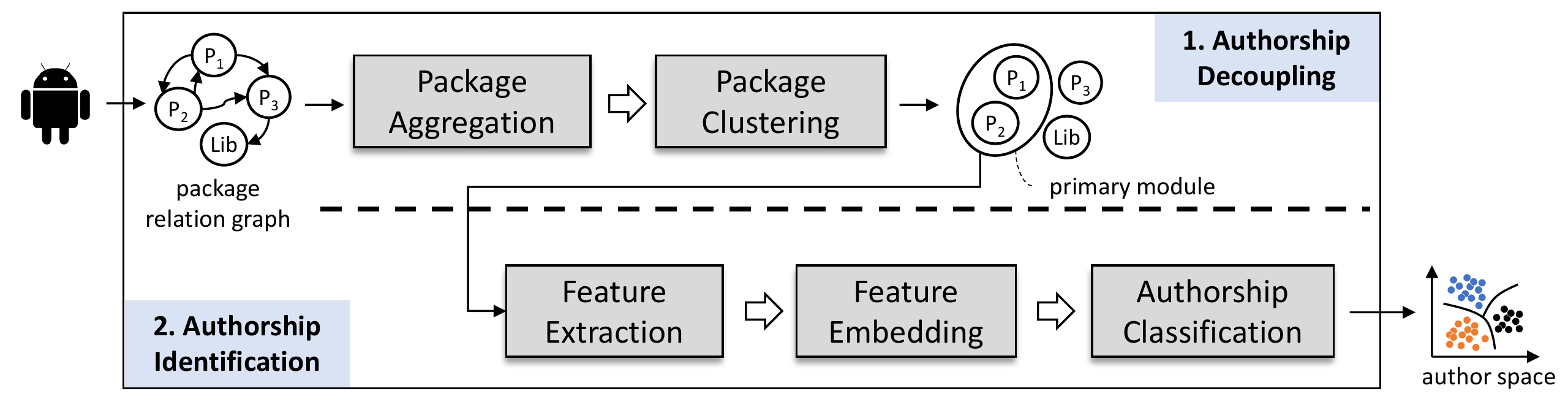}
	\caption{System overview of \tool}
	\label{fig:overwiew}
\end{figure*}
\subsection{Component Analysis}

A large number of apps have structural complexity and it is not realistic to analyze all the code of apk for authorship attribution. Besides, due to multi-module development, the code in an app may be attributed to an individual or a team with cooperation. 
Each team develops its own module code, the coupling between modules is poor, and the aggregation within modules is high. 
Even worse, it is observed that the proportion of code written by the author is small. 
As shown in Fig.~\ref{fig:codepercent}, among the 100 apps randomly selected from F-Droid, this proportion is less than 30\% in about 90\% of apps. Due to some limitations, \textsc{LibScout}~\cite{libscout} and \textsc{LibD}~\cite{libd} can only detect a subset of third-party libraries from apps. 
Consequently, it is non-trivial to classify a mix of code to specific authors. Prior studies~\cite{authorship2018codaspy,Kalgutkarauthorship} fail to consider this phenomenon in their approaches although they also achieve good results. 

Motivated by the above reasons, our goal is to develop a new authorship identification approach. Before authorship identification, we establish the correlation between packages for the app analyzed, and divide packages into independent modules. 
Among these modules, there is a module that includes \textsf{\small MainActivity} and it can be easily identified by querying \textsf{\small AndroidManifet.xml}. As the entry point of activities, \textsf{\small MainActivity} is the core class of an Android app and implements the main functions by means of function calls, ICCs, etc. Therefore, according to \cite{PiggyApp}, the module where this activity resides is regarded as the primary module designed by the leading author and can be used for authorship identification.

\begin{table}
	\centering
	\caption{The result of granularity comparison experiment}
	\label{tab:granularity}
	\begin{tabular}{cccccc}\toprule
		\textbf{Granularity} & \textbf{Precision} & \textbf{Recall} & \textbf{F1-score} & \textbf{Accuracy}& \textbf{Time(s)}\\ \midrule
		Class & 94.43\% & 97.01\% & 94.43\% & 94.94\% & 5113.29\\
		Package & 96.00\% & 97.69\% & 95.99\% & 95.81\% & 1357.46\\
		\bottomrule
	\end{tabular}
\end{table}
\subsection{System Overview}

We propose a framework \tool with two major phases to identify the primary module of given apps and generate the representations with regard to authors' programming styles for authorship attribution. Fig.~\ref{fig:overwiew} presents the overview of our model. Given an app, it proceeds as follows:
\begin{itemize}[leftmargin=*]
	\item \textbf{Authorship decoupling}: We treat Java packages as units of analysis and divide them into different modules for each app. 
	We create a package relation graph for a given app in terms of call relationship, inheritance relation, and ICC connection. We employ \emph{package aggregation} to group all packages created by the authors and design semantic and structural similarities to calculate the weight of each pair of them. Last, we utilize the Louvain model and associated weights between packages to divide packages into modules. The primary module can be determined according to where \textsf{\small 
MainActivity} is located.  
	\item \textbf{Authorship identification}: We extract three types of features from the primary module of an app and make use of the TF-IDF algorithm to identify the important sequences of these features. Then, we form feature vectors based on the \emph{word2vec} model and select three machine learning models as supervised classifiers to predict a potential author. 
\end{itemize}

\noindent\textbf{Granularity of package}. In this study, we take \emph{packages} rather than \emph{classes} as atomic units of module since classes under the same package, especially with multiple levels, are likely written by the same author. To prove this hypothesis, we extracted 100 authors from the F-Droid dataset and randomly selected one app from each of them. Then, according to the method in Section~\ref{sec:apkfilter}, we carry out the authorship decoupling with class granularity and that with package granularity. Table~\ref{tab:granularity} shows the comparison result, and it is observed that authorship decoupling in the granularity of package achieves higher recall, accuracy, and F1-score than that with class granularity. Moreover, it costs only one-fourth time. 
\section{Authorship Decoupling}\label{sec:apkfilter}

An Android app is composed of several functionality-independent modules~\cite{PiggyApp}. 
Similarly, the app oftentimes integrates diverse code from multiple developers~\cite{binauthor2019jcs,authorship2019csur}, including third-party libraries, the integrity of Android SDKs, and repackaged code. 
Therefore, we propose authorship decoupling to attribute part of code to specific developers.

\subsection{Package Relation Graph}\label{sec:decoupling:overview}

Different from functionality decoupling, authorship decoupling clusters code in terms of authors rather than functionality. 
Since one author can develop several functions that may not have explicit semantic relationships (\eg, call relationship) in between, we propose several coarse-grained strategies to distinguish code developed by varying authors. 
Without loss of generality, we propose a package relation graph to represent an Android app as follows.

\begin{mydef}%
	An Android app can be represented as a weighted directed multigraph $G = (V, E, \phi)$, where $V$ is the set of packages in the app, $E$ is the set of edges between $V$, and $\phi$ is authorship function that $\phi(u)$ represents the author for package $u$. 
\end{mydef}

In the graph $G$, we use an edge $e$, where $e = (u, v) \wedge u, v\in V$, to express  the semantic relationships between two packages. Here we consider three types of semantic relationships: \emph{call relationship}. If there is an invocation from one construct (\eg, method or variable) in package $u$ to another method or variable in package $v$, we treat that there is a call relationship between $u$ and $v$; \emph{inheritance relationship}. If one class in package $u$ is inherited from another class in package $v$, it implies that $u$ and $v$ have an inheritance relationship; and \emph{ICC connection} reveals a type of connection between two packages $u$ and $v$ if one component class in $u$ has an Inter-Component Communication with another class in $v$. 

\subsection{Package Aggregation}\label{aggregation}

To group all packages created by the same authors, we make three assumptions in consideration of app development. 
First, packages defined in known third-party libraries are regarded as being owned by one author. It is rational since third-party libraries are developed by single organizations and should be separated from the primary module in apps. 
Second, components defined in \textsf{\small AndroidManifest} are probably involved with one author. \textsf{\small AndroidManifest} defines all essential information about an app, of which the components are Activity, Service, Broadcast Receiver, and Content Provider for underlying apps' functionality. As such, these components are created by the same author of high probability. We investigated 100 apps from dataset F-Droid and found that 89\% of apps had components in the same module. 
Third, vertices in a circle are probably created by one single author~\cite{tang2019cs}. Authors in one app have asymmetric knowledge, as the caller of third-party libraries is aware of all exposed interfaces by libraries, however, the callee does not know the caller's interfaces at all. Therefore, if there are bidirectional relations between vertices $u$ and $v$ (\ie, $u$ and $v$ are in a circle), it is most likely that $u$ and $v$ are created by the same author. 
We take an app called ``\textsf{\small MinCal Widget}" as an example. Its package structure is shown in Fig.~\ref{fig:package}. We filter out the third-party packages ``\textsf{\small Landroid/*}'' and ``\textsf{\small Landroidx/*}'', and generate a directed graph $G$ with the remaining ones. Then, the components declared in \textsf{\small AndroidManifest} belong to the same author. All the components of this app are located in the package``\textsf{\small minimalcalendarwidget}'' and ``\textsf{\small activity}'', so the two packages belong to the same author. Then, we look for circles and find that packages ``\textsf{\small minimalcalendarwidget}'', ``\textsf{\small activity}'', ``\textsf{\small external}'', ``\textsf{\small receiver}'', ``\textsf{\small service}'' are in the same circle. As such, the four packages are generated by the same author.

Based on the above assumptions, we perform an algorithm (as Algorithm~\ref{alaggregation}) to aggregate packages by the same authors. 
In the beginning, we assign a unique id to each package as a distinguishable author (lines 1-2). We recognize all contained libraries in this app (line 3), and treat the libraries within the same author (lines 4-6). 
We retrieve all the components defined in \textsf{\small AndroidManifest}, and re-assign the same author id to these packages from line 7 to 9. 
Then we take advantage of the depth-first search algorithm to find all circles in the graph $G$ at line 10-11. 
More specifically, for the last node $v_n$ passed to the function \textsc{\textsf{\small DFS\_Circle\_Detect}}, we traverse all the outgoing edges to examine whether there is a backtracking edge to previously visited nodes. A circle is detected if found (line 14), and we determine that all the packages in a circle are created by the same author. 
Otherwise, a recursive process is performed at line 18. 
At last, $\phi$ is updated for the app.

\begin{algorithm}[t]
	\caption{Package Aggregation}\label{alaggregation}
		\KwIn{ $G = (V, E, \phi)$: an Android app }
		\KwOut{ $\phi$: authorship function }
		\For {$u$ in V}{
		 	$\phi(u)$ = uniq\_id;  //assign unique author id to nodes \\
		}
		libs $\subset~V$ $\gets$ identify contained libraries; \\
		\For {$lib$ in $libs$}{
		    \For {$u, v$ in $lib$}{
		        $\phi(u) = \phi(v)$; \\
		    }
		}
		$comps \gets $ retrieve components in AndroidManifest; \\
		\For {$u, v$ in $comps$}{
		    $\phi(u) = \phi(v)$;
		}
		\For{$u$ in $V$}{
		    \textsc{DFS\_Circle\_Detect}($\langle u \rangle$); \\
		}

		\SetKwProg{DFS}{Function}{:}{}
		\DFS{\textsc{DFS\_Circle\_Detect}($\langle v_0, v_1,...v_n\rangle$)}{
		
			\For {$u$ in $\{u | (u, v_n) \in E\}$}{
			    \If{$\phi(u)\neq\phi(v_n)~\wedge~u \in \langle v_0, v_1,...v_{n-1}\rangle$}{ 
			        \For{$v_i$ in $\langle v_t,...v_n\rangle$ where $v_t=u$}{
			            $\phi(v_i) = \phi(u)$; \\
			        }
			    }\ElseIf{$\phi(u)\neq\phi(v_n)$}{
			        \textsc{DFS\_Circle\_Detect}($\langle v_0, v_1,...v_n, u \rangle$); \\
			    }
		 	}
	 	}
		
\end{algorithm}

\subsection{Package Clustering}\label{relatcorr}

Part of the same authors' packages have been found in the previous steps. But there are still some packages that cannot be glued with the assumptions. Hence, we cluster packages in this section to cope with the rest. To this end, we develop two weights for distance between packages in the graph. 

\noindent\textbf{Semantic Distance. } 
As there are three types of semantics (\ie, call relationship, inheritance relationship and ICC connection) between two packages. Given an edge $(u,v)~\wedge~\phi(u)\neq~\phi(v)$, we use $n^c_{uv}$, $n^h_{uv}$ and $n^i_{uv}$ to represent the number of invocations, inheritance cases, and ICC links, respectively. 
Hence, the semantic distance can be computed as:
\begin{flalign}
dist(u, v) = \frac{1}{n_{uv}^{c}+n_{uv}^{h}+n_{uv}^{i}}
\label{eqt:semantic}
\end{flalign}
Noted that, $dist(u, v)$ is assigned with the maximal value between $u$ and $v$. For a relation graph, we employ Floyd's algorithm~\cite{floyd} to calculate the shortest path between every two packages and then update $dist(u,v)$ with their Floyd distance. 
Based on that, we propose the following to represent the correlation between $u$ and $v$.
\begin{flalign}
corr(u,v)=e^{-min\left (dist\left ( u,v\right ),dist\left ( v,u\right )\right )}
\label{eqt:dis}
\end{flalign}
The larger the value of $corr(u,v)$ is, the greater the correlation between package $u$ and package $v$.

\begin{figure*}[!t]
	\begin{minipage}[t]{0.31\linewidth}
		\centering
		\includegraphics[width=1\textwidth]{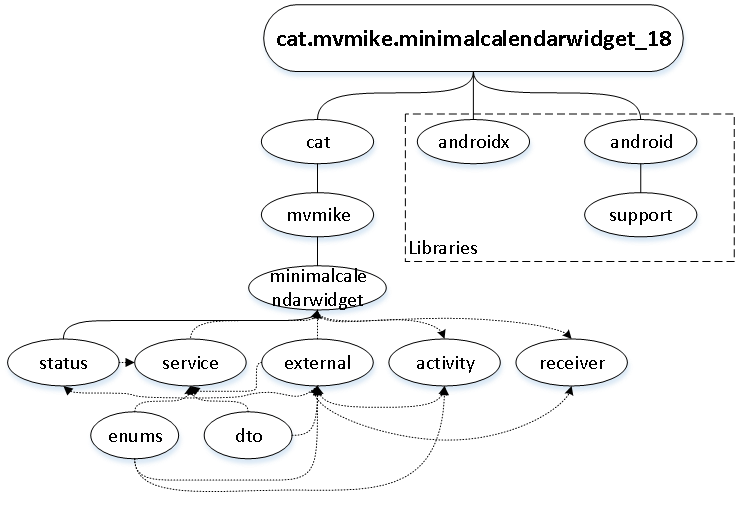}
		\caption{An illustrative example for package structure of one app} 	
		\label{fig:package}
	\end{minipage}
	\hfill
	\begin{minipage}[t]{0.31\linewidth}
		\centering
		\includegraphics[width=1\textwidth]{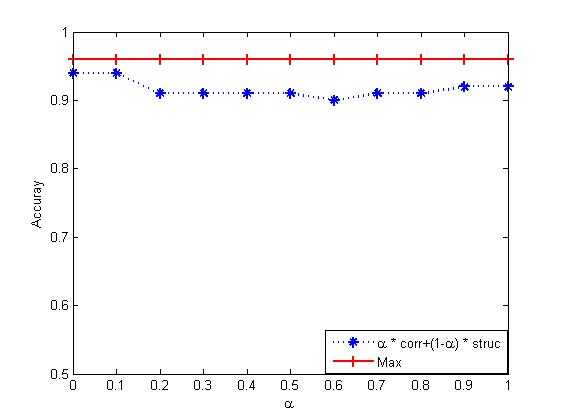}
		\caption{Different experimental accuracy of $sim(u,v)$ for authorship decoupling} 
		\label{fig:weight}
	\end{minipage}%
	\hfill
	\begin{minipage}[t]{0.31\linewidth}
		\centering
		\includegraphics[width=1\textwidth]{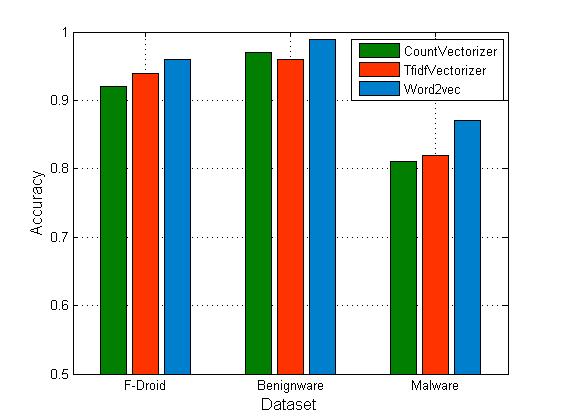}
		\caption{Detection accuracy for authorship identification, under three vectorization methods}
		\label{fig:word2vec}
	\end{minipage}
\end{figure*}

\noindent\textbf{Structural Distance.} 
Package structure can also aid in authorship decoupling. As shown in Fig.~\ref{fig:package}, there are two packages ``\textsf{\small cat.mvmike.minimalcalendarwidget.external}'' and ``\textsf{\small cat.mvmike.minimalcalendarwidget.status}''. They are likely created by the same author considering the long common prefix of their names. 
Therefore, we put forward structural features for authorship decoupling. 
We calculate the structural similarity based on the nearest common parent (NCP) between every two packages. Given two packages $u$ and $v$, their structural similarity can be computed as:
\begin{flalign}
struc(u,v)=\sum_{i=1}^{n}\frac{1}{i}
\label{eqt:sim}
\end{flalign}
Where $n$ is the depth of the NCP of package $u$ and $v$. The app name is defined as the root node and its depth is one. Noted that if $u$ is the parent of $v$, the NCP of $u$ and $v$ is $u$.

According to the semantic and structural similarities, we compute the normalized probability of package $u$ and $v$ being under the same author as: 
\begin{flalign}
sim(u, v) = max(nor(corr(u,v)), nor(struc(u,v))) \label{eqt:weight} 
\end{flalign}
where $nor(\cdot)$ is a normalization function via \emph{Min-Max Feature Scaling}. The degree of association between packages can be computed via $sim(u,v)$. 
To evaluate its effectiveness, we propose another combination method as $\alpha*corr+(1-\alpha)*struc$, where $\alpha$ represents the proportion of two similarities. We randomly selected 100 apps from dataset F-Droid and measured the accuracy in authorship decoupling, respectively. The comparison result is shown in Fig~\ref{fig:weight}, and it is seen that using the maximal value of these metrics works better than the parameterized.

At this point, we get the correlation weight between packages in a given app. Given two packages $u$ and $v$, if $\phi(v)=\phi(u)$, the weight between $u$ and $v$ is one, otherwise, the weight between $u$ and $v$ is the value of $sim(u,v)$. Then, we use the Louvain~\cite{louvain} model to represent the tightness between packages in an app, and divide the packages into different clusters. Taking packages and their associated weights as input, Louvain outputs the module to which each package belongs.
After that, we select the primary module where the activity is located as the module designed by the leading author, and then identify the code authorship of the primary module. 

\section{Authorship Identification}\label{sec:authorshipIdentification}
In this section, we utilize the authors' distinctive writing styles to identify the most likely author for a particular workpiece from a group of candidates. 
We describe how to extract features, embed features, and make predictions. 
\subsection{Feature Extraction}
Feature extraction is a key part of authorship identification, and its goal is to extract features with author's distinct writing style. 
Instead of traditional code authorship attribution, we should not only extract features from \textsf{\small .dex} file and consider other resource files, \eg, \textsf{\small AndroidManifest.xml}. Note that we only extract the source code features from the primary module identified in Section~\ref{sec:apkfilter} and exclude the influence of Android APIs.  
In order to better reflect the characteristics of features as fingerprints, the features we analyzed are described as follows: 
\begin{table}
	\centering
	\caption{Summary of different types of stylometric features}
	\label{tab:features}
	\begin{tabular}{ccp{4.6cm}}\toprule
		\textbf{Category} & \textbf{Name} & \textbf{Description}\\ \midrule
		\multirow{3}{*}{Dex}
		&Identifier Name & Identifiers of classes, fields, and methods\\
		&API Calls & The sequence of APIs invoked in a class\\
		&Instructions & Instruction sequences from \textsf{\small .dex} file \\ \midrule
		Lib & Library name & The name of third-party libraries \\ 
		\midrule
		\multirow{4}{*}{Manifest}&
		component & Naming rules of activities, services, providers, and receivers \\
		&uses-feature &  External hardware/software features \\
		\bottomrule
	\end{tabular}
\end{table}
\begin{itemize}[leftmargin=*]
	\item \textbf{Dex-level features}. The \textsf{\small .dex} file contains all the Java/Kotlin code of an app, and we extract three types of string features: \emph{identifier names}, including class name, method name, and field name. To eliminate the influence of the Android framework, we do not take into account overloaded methods, onCreate and onPause; \emph{instruction sequence}, that is how authors write code to implement their functionality. It is worth mentioning that this is literal sequences while not sequences in a control flow graph, as they preserve authors' programming habits; \emph{the use of Android APIs}, which reveals to some extent authors' familiarity and preferences in Android development. For example, to build an HTTP connection, authors may use \textsf{\small HttpURLConnection} or \textsf{\small Apache HttpClient} offered by \emph{Apache}.
	\item \textbf{Manifest-level features}. Each app contains an \textsf{\small AndroidManifest.xml} file, which defines the package name, components, permissions, etc. Different components serve for different purposes. The uses-feature declares external hardware or software features at runtime. These configurations are related to the functionality of the app, and authors tend to reuse the same setting when publishing the same type of apps. Hence, the naming of each component is related to the author's naming rules and functionality provided. Here we extract the names of activities, providers, services, broadcast receivers, and use-features from \textsf{\small AndroidManifest.xml}.
	\item \textbf{Lib-level features}. Third-party libraries are required to perform additional functions at runtime. Generally, there are several candidates for the desired function. For example, developers can integrate either \emph{AdMob} or \emph{Unity Ads} for advertisement, and either \emph{Google Analytics} or \emph{Crashlytics} for diagnosing crashes of apps. The selection of third-party libraries can reveal the habits and personality of app authors.
\end{itemize}

 Table~\ref{tab:features} summaries the features used for authorship identification. These features contain a lot of noise and are not suitable for analysis of large datasets. Hence, we use the TFIDFVectorizer~\cite{datamining} to extract key sequences of word-level strings from string features. TFIDFVectorizer constructs n-grams features from three sources, \ie, API calls, identifier names, and instructions. 
 During the dex-level feature extraction process, we extract the three features in turn and use it to generate high-frequency sequences for vectorization. 
 Then we tune proper hyper-parameters for TFIDFVectorizer. We set \emph{max\_features} as 50, \emph{min\_df} as 3, \emph{n-grams} range from 3 to 5. This provides the best trade-off between accuracy and processing time.

\subsection{Feature Embedding}
\tool uses the \emph{word2vec} model to process text vectorization for extracted features. 
As a distributed representation, \emph{word2vec} uses low-dimensional dense vectors to represent the semantic information of words through text learning, which is a good measure to measure the similarity between words. 
To verify its effectiveness, we compare the accuracy of CountVectorizer~\cite{datamining}, TFIDFVectorizer, and \emph{word2vec} on the experimental results. These three methods are common methods of text vectorization. Among them, CountVectorizer converts the words into a word frequency vector, and then counts the number of times each word appears. TFIDFVectorizer converts the text into \emph{tf-idf} feature vector. Fig.~\ref{fig:word2vec} shows the results of these three embedded techniques. In this experiment, we select F-Droid, benignware, and 2,900 malware apps as datasets, and extracted the same number of \emph{top-k} features. We found that the performance of \emph{word2vec} was a litter higher than the other two methods. In consequence, we choose \emph{word2vec} for feature embedding in this study. At last, 
in order to improve the accuracy of prediction, we set \emph{windows} as 3, \emph{min\_count} as 10, and set the maximum number of columns of the feature vector to 1,000.

\subsection{Authorship Classification}
The generated developers' fingerprints are used for the author prediction. The feature vectors generated in the previous stage as input data, the author attribution problem is seen as a supervised learning classification problem, which classifies unknown apps to their corresponding developers. As a comparison, we also investigate and build three typical supervised machining learning models for our classification tasks and evaluate their effectiveness. As our problem is a typical multi-classification problem, we choose support vector machine (Linear SVM), random forest, and logistic regression to measure the performance of each machine learning classifier.
 
\section{Evaluation}\label{sec:eval}
In this section, we first propose four research questions to answer, and then introduce the experiment setting. 
We aim to address the following questions:

\begin{enumerate}[label=\textbf{RQ\arabic*.},leftmargin=*]
	\item How effective is authorship decoupling, and what extra-essential code resides in an apk file (see Section~\ref{sec:eval:module})?
	\item How effective of \tool in authorship identification, and what is the improvement by authorship decoupling (see Section~\ref{sec:eval:authorship})?
	\item How resilient is \tool to obfuscation (see Section~\ref{sec:eval:obfuscated})?
	\item Compared with \textsc{AppAuth}, how effective is \tool in identifying the author of a given app (see Section~\ref{sec:eval:compared})?
\end{enumerate}

\subsection{Experimental Setup}\label{sec:eval:setup}

\noindent\textbf{Dataset}. 
To evaluate our methodology, we collected a total of 32,968 Android apps from various sources as shown in Table~\ref{tab:data}. 
The following describes how we collect these apps:

\begin{itemize}[leftmargin=*]
	\item \textbf{F-Droid}. It is an open-source repository for free Android apps~\cite{fdroid}. To date, there are thousands of apps maintained by F-Droid as well as their source code. We obtain 2,296 apps with source code via its API in total.
	\item \textbf{Benignware}. We obtained 1,672 whitelisted apps from ANVA~\cite{anva}. ANVA is an industry alliance that is responsible for monitoring, detecting, and responding to network threats. Every year, it publishes a white list of Android apps that are thoroughly examined by 11 professional security assessment institutes. 
	\item \textbf{Malware}. We collect 29,000 Android apps from the wild (including Google Play~\cite{googleplay}, ApkPure~\cite{apkpure}, Anzhi~\cite{anzhi}, etc) which are labeled by \textsc{VirusTotal}~\cite{virustotal} as malware. 
	\item \textbf{Obfuscated Apps}. We select a number of malware samples and obfuscate 10\% of them per author as a test set via \textsc{ProGuard}~\cite{proguard}. \textsc{ProGuard} is used for string obfuscation and code shrinking. This dataset is used to evaluate how resilient our approach is to obfuscation. 
\end{itemize}

As not all authors create plenty of apps (some of them only create one app), it raises the difficulty of fingerprinting their programming styles. Therefore, we employ a \emph{least-apps} policy by which we only retain the authors having at least $n$ apps. 
We set $n$ with 3 for dataset F-Droid considering the relatively low app number owned by one author. For the other three datasets, $n$ is set to 10. 
In addition, to enable a more accurate evaluation, we discarded 8,691 apps signed by the public certificates~\cite{sign2019ase} and duplicate apps published in multiple markets. Hence, we obtain 492 apps from 164 authors for dataset F-Droid, 686 apps from 17 authors for dataset Benignware, and 14,564 apps from 100 authors for dataset malware. Obfuscated apps are created by sampling 2,900 malware apps, each author contains 100 apps.

\noindent\textbf{Implementation}. We implement \tool on top of several state-of-the-art tools. The decompilation of Android apps mainly relies on \textsc{AndroGuard}, for extracting authorship features, call relations, and inheritance relations. ICC connections are extracted via \textsc{IC3-DialDroid}~\cite{ic3-dialdroid}, and we re-allocate the connections from Java classes to their belonging packages. The classification task is fulfilled by building applications on top of Python library \textsc{Sklearn}~\cite{sklearn} and \textsc{Gensim}~\cite{gensim}. 
All the evaluation experiments were conducted on the environment  with 8-core i7 Intel CPU and 12GB of RAM.

\noindent\textbf{Model configuration and metrics.} For different research tasks, we use precision, recall, F1-score, and accuracy as metrics for comparison according to~\cite{researchmethod}. 
From RQ2 to RQ4, we group authors based on certificates and generate true results. We employ \tool to generate predicted results. We use 90\% of the data as the training set and the remaining 10\% as the test set. For every experiment, we perform 10-fold cross validation.

 \begin{table}
	\centering
	\caption{Statistics of Experimental Datasets} 	\label{tab:data}
	\begin{tabular}{crccc}\toprule
		\multirow{2}{*}{\textbf{Dataset}} & \multirow{2}{*}{\textbf{Total}} & \multicolumn{3}{c}{\textbf{Least-Apps Filtering}} \\ \cline{3-5}
		& & \textbf{\# Least Apps} & \textbf{Authors} & \textbf{Apps} \\ \midrule
		F-Droid & 2,296 & 3 & 140 & 416 \\ 
		Benignware & 1,672 & 10 & 17 & 686\\
		Malware & 29,000 & 10 & 100 & 14,564 \\
		Obfuscated & 2,900 & 100 & 29 & 2,900 \\
		\bottomrule
	\end{tabular}
\end{table}

\begin{table}
	\centering
	\caption{Effectiveness of Authorship decoupling on F-Droid apps}
	\label{tab:moduleresult}
	\begin{tabular}{ccccc}\toprule
		\textbf{Sample} & \textbf{Accuracy} & \textbf{Precision} & \textbf{Recall} & \textbf{F1-score}\\ \midrule
		\tool & 96.11\% & 97.35\% & 96.11\% & 95.70\%\\
		\textsc{PiggyApp} & 81.18\% & 86.01\% & 81.87\% & 76.77\% \\
		\bottomrule
	\end{tabular}
\end{table}

\subsection{Authorship Decoupling Evaluation (RQ1)}\label{sec:eval:module}

As an apk file contains code from one or more authors, we propose an authorship decoupling technique to distinguish the primary code from the others. To evaluate its effectiveness, we first build the ground truth from dataset F-Droid, and then examine how accurately our approach decouples the authorship of code. 

\noindent\textbf{Building the ground truth. }
Since F-Droid hosts thousands of free Android apps and the corresponding source code, we retrieve the files in source code to establish a ground truth for authorship decoupling evaluation. In the ground truth, all the classes of a given app are divided into two parts, \ie, the primary module generated by the leading author and the non-primary module that contains the remaining classes. 
\begin{enumerate}[leftmargin=*,label=Step \arabic*]
    \item We first use \textsc{Androguard} to decompile an APK file to get its classes and \textsf{\small MainActivity}. Note that we do not consider inner classes, \eg, \textsf{\small Test\$1.class}.
    \item We take advantage of ``\textsf{\small settings.gradle}'' to determine the primary module where \textsf{\small MainActivity} resides, and then we get the classes contained in the primary module. In an Android project, \textsf{\small settings.gradle} is a configuration file for subprojects, whose goal is to manage multiple modules. Each module name corresponds to its root directory. We can walk through all the \textsf{\small .java} and \textsf{\small .kt} files to determine which module they belong to. 
    \item Last, we classify the classes that do not appear in the primary module as the non-primary module. Besides, we check whether any classes are not compiled into the \emph{.dex} file classes. This is reasonable because some test code has been retained in the published source code and it could negatively influence the reliability of our ground truth. 
\end{enumerate}

Based on the above method, we build the ground truth of 416 F-Droid apps for authorship decoupling. 

\noindent\textbf{Result analysis. }After authorship decoupling, we get the primary module and classes it contains. The remaining classes are merged into the non-primary module. Therefore, it can be 
treated as a binary classification for a given app. We use metrics (\eg, precision) to evaluate the package decoupling performance of each app, and then calculate the average of all apps as average metrics. 
Moreover, we re-implement \textsc{PiggyApp}~\cite{PiggyApp}, which provides a method for module decoupling based on class inheritance, method calls, and etc. 
The comparison result is shown in Table~\ref{tab:moduleresult}. 
Compared to \textsc{PiggyApp}, our model gets much better results, \eg, accuracy is increased by 14.93\% and F1-score by 18.93\%. 
The improvement stems from higher quality features employed in this study, particularly the structural and semantic similarities. These features prove to be effective in module decoupling. In addition to authorship attribution, authorship decoupling can be used for code plagiarism, functionalities classification, and app version detection, with reducing noise interference.

In addition, on the premise of excluding the interference of inner class, we compare the classes of the same package in the source code and the apk file, and note that the classes of the APK package do not all exist in the source code. These classes are from four sources as follows:
\begin{itemize}[leftmargin=*]
	\item \textbf{Auto generation}. During compilation, an Android app automatically generates classes \textsf{\small R}, \textsf{\small BR} and \textsf{\small BuildConfig}. Files \textsf{\small R}, and \textsf{\small BR} record the resource information. The \textsf{\small BuildConfig} file records the configuration information of \textsf{\small build.gradle}.
	\item \textbf{Lambda expression}. Java 8 introduces $Lambda$ expressions, a compact way to represent behaviors. Its use reduces the template code and makes the data flow processing logic clear. These generated \textsf{\small .class} files are found in the apk file, but not exist in the source code. For example, when analyzing the app ``\textsf{\small app.fedilab.nitterizeme\_9}'', we found a series of classes like ``\textsf{\small Lapp/fedilab/lite/helper/-\$\$Lambda...}''.
	\item \textbf{Classes generated by the author instructions}. When writing code, developers may define a new class or upload classes to specified packages from the Internet in the statement, \eg, new STTFragment(). These files may be irrelevant to developers.
	\item \textbf{Different modules with the same package}. It is possible that the same package may contain different module classes. Taking the app ``\textsf{\small com.termoneplus\_324}'' for example, it contains three modules--$samples$, $term$ and $libtermexec$. However, modules $term$ and $libtermexec$ have the same package ``\textsf{\small Lcom/termoneplus}''. When the Android project is compiled into an APK, the classes under the package of these two modules are packaged into the same package, although they do not belong to the same module. 
\end{itemize}

\begin{figure}
	\begin{minipage}[t]{0.49\linewidth}
		\centering
		\includegraphics[width=1.0\textwidth]{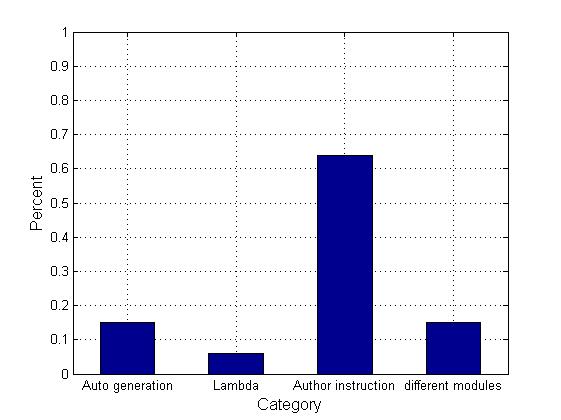}
		\caption{The percentage of each type in the total errors in authorship decoupling}
		\label{fig:factors}
	\end{minipage}%
	\hfill
	\begin{minipage}[t]{0.49\linewidth}
		\centering
		\includegraphics[width=1\textwidth]{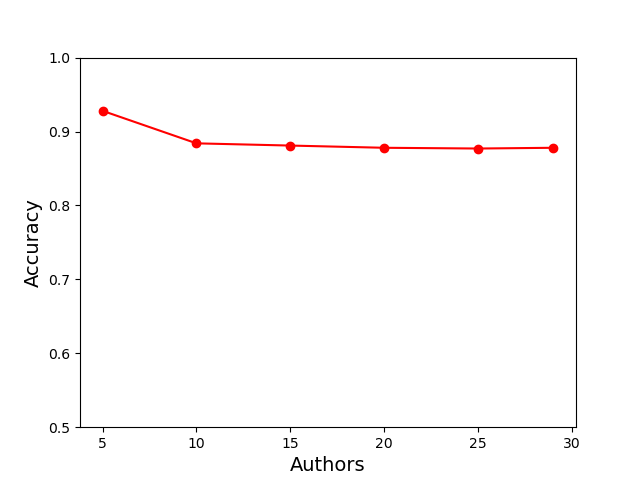}
		\caption{The accuracy of authorship identification under different number of authors}
		\label{fig:authornumber}
	\end{minipage}
\end{figure}

Fig.~\ref{fig:factors} depicts the proportion of the four causes mentioned above. The classes generated by the author's instructions are the most important factor of authorship decoupling error, accounting for 64\%. 
In the future study, we should eliminate the noise effect of these four classes on authorship identification.

\subsection{Authorship Identification Evaluation (RQ2)}\label{sec:eval:authorship}

\begin{table*}[t]
	\small
	\centering
	\caption{authorship identification ACCURACY FOR different DATASETs}\label{tab:classifierresult}
	\begin{tabular}{ccccccc} \toprule
		\multirow{2}{*}{\textbf{Classifier}} & \multicolumn{2}{c}{\textbf{F-Droid}} & \multicolumn{2}{c}{\textbf{Benignware}} & \multicolumn{2}{c}{\textbf{Malware}}  \\ \cline{2-7}
		& \textbf{Accuracy} & \textbf{Runtime(s)} & \textbf{Accuracy} & \textbf{Runtime(s)} & \textbf{Accuracy} & \textbf{Runtime(s)} \\ \midrule
		Linear SVM & 96.5\% & 6.33 & 98.9\% & 1.62 & 83.1\% & 277.0 \\
		Random forest & 95.7\% & 1.44 & 99.0\% & 0.23 & 82.4\% & 35.56 \\
		Logistic regression & 95.8\% & 1.78 & 98.8\% & 0.18 & 82.6\% & 6.94\\
		\bottomrule
	\end{tabular}
\end{table*}

\begin{table}[t]
	\small
	\centering
	\caption{Accuracy of Authorship Identification in three datasets.}\label{tab:compareresult}
	\begin{tabular}{ccccccc} \toprule
		\textbf{Method} & \textbf{F-Droid} & \textbf{Benignware} & \textbf{Malware}  \\ \midrule
		Primary code & 95.7\% & 99.0\% & 87.4\% \\
		Whole apk & 91.1\% & 97.8\% & 86.0\% \\
		\bottomrule
	\end{tabular}
\end{table}

In this section, we evaluate whether \tool is effective in identifying the author of a specific piece of code. 

\noindent\textbf{Result Analysis.} We use F-Droid, benignware, and malware datasets to evaluate the accuracy of authorship identification. As shown in Table~\ref{tab:classifierresult}, for these three datasets, we get the accuracy of around 96\%, 98.9\%, 82.7\%, respectively. Overall, our model provides high accuracy results in authorship identification. We also compare the performance of the three supervised machine learning models in authorship identification. We find that the accuracy of the three classifiers is similar. Linear SVM has gained a narrow lead in average accuracy but runs far longer than the other two classifiers. Random forest and logistic regression are more suitable for our model in terms of running time and effectiveness. 
Besides, when reviewing the package names and versions of apps in dataset F-Droid, we notice that the result is more like the detection of different versions of the same app. The considerably high accuracy enables \tool to be efficient in detecting highly similar code such as plagiarism detection.

In order to evaluate the model more comprehensively, we select random forest to compare the performance of authorship identification based on the primary module and the whole apk (without authorship decoupling). Table~\ref{tab:compareresult} describes the results of the comparison. \tool achieves the accuracy of 95.7\%, 99.0\%, and 87.4\% for datasets F-Droid, benignware, and malware samples, respectively. 
It is observed that authorship decoupling makes an increase of 4.6\% in accuracy on dataset F-Droid, but only makes an improvement of 1.2\% and 1.4\% for benignware and 2,900 malware samples. The similar accuracy may be attributed that there are a number of apps created by the same author for only one module without third-party libraries. Besides, the result can also be caused by code clone among apps. 
We have investigated 10 authors that create at least 10 apps, and 12\% of apps are found with a similar code structure. 
Besides, we further evaluate the impact of the number of authors on authorship identification. With this purpose, we used the obfuscated dataset with 29 authors. Each author has the same number of apps, and this excludes the interference of different authors having different numbers of apps to the result. The experimental result is shown in Fig.~\ref{fig:authornumber}. We find that the accuracy of authorship decreases slightly with the increase of author number. Overall, accuracy does not change significantly. This result shows that our model is not disturbed by the number of authors, and has strong stability. 
\subsection{Obfuscation-resilient Analysis (RQ3)}\label{sec:eval:obfuscated}

\begin{table}
	\centering
	\caption{Evaluation on Obfuscation Resiliency in Authorship Identification.}
	\label{tab:obfuscated}
	\begin{tabular}{ccccc}\toprule
		\textbf{Sample} & \textbf{Precision} & \textbf{Recall} & \textbf{F1-score} & \textbf{Accuracy}\\ \midrule
		Non-obfuscated & 85.0\% & 86.0\% & 85.0\% & 87.6\%\\
		obfuscated & 80.4\% & 81.4\% & 80.2\% & 80.4\%\\
		\bottomrule
	\end{tabular}
\end{table}
\begin{table}[t]
	\centering
	\caption{Authorship Identification compared with \textsc{AppAuth}.}
	\label{tab:compared}
	\begin{tabular}{ccccc}\toprule
		\textbf{Sample} & \textbf{Precision} & \textbf{Recall} & \textbf{F1-score} & \textbf{Accuracy}\\ \midrule
		\tool & 88.4\% & 87.6\% & 87.1\% & 87.8\%\\
		\textsc{AppAuth} & 86.4\% & 84.5\% & 84.2\% & 84.4\%\\
		\bottomrule
	\end{tabular}
\end{table}

In this section, we evaluate whether \tool is resilient to obfuscation in Android apps. Considering the effectiveness and running time, we choose the random forest for the comparative experiment on the obfuscated dataset. For this dataset, we randomly select 90\% as the training set and the remaining 10\% as the test set for each author. For the test set, we use \textsc{ProGuard} to obfuscate apps. There are three steps for \textsc{ProGuard} to obfuscate Android apps. Initially, unused code (\eg, unused methods and variables) is removed to shrink the size of code. Then, Java bytecode optimization is performed. Finally, it performs name obfuscation, \ie, it obfuscates code by renaming variables, classes, and methods into short, random, meaningless words. Note that we do not perform obfuscation on Android APIs. 
After obfuscation, we take the obfuscated apps and their non-obfuscated apps as the test set, respectively, and measure the resulting accuracy of authorship identification. 

\noindent\textbf{Result analysis.}  Table~\ref{tab:obfuscated} shows the experimental results of the obfuscated and non-obfuscated test sets, with the accuracy of 87.6 and 80.4\%, respectively. Compared with the non-obfuscated apps, the accuracy of the obfuscated apps decreases slightly, but not significantly. That's because manifest-level features are not obfuscated. Moreover, we investigate five apps and their obfuscated apps. We compare instruction sequences of each pair of apps and find that they are less affected by code obfuscation. The result suggests that \tool is somewhat resistant to code obfuscation. 
\subsection{Comparison with \textsc{AppAuth} (RQ4)}\label{sec:eval:compared}
To validate the performance of \tool more fully, we compare it with \textsc{AppAuth} in terms of authorship attribution. 
\textsc{AppAuth} is an Android authorship detector and a novel learning-based approach for predicting the authorship of app clones. We select a dataset of 2,900 non-obfuscated apps and then use the two tools in turn. 
 Table~\ref{tab:compared} presents the performance of both \textsc{AppAuth} and \tool. Compared with \textsc{AppAuth}, our approach raises a 3.4\% improvement to identification with the accuracy of 87.8\%. 
Therefore, our two-phased approach proves to be effective in Android authorship attribution.

\section{Threats to Validity}\label{sec:discuz}

\noindent\textbf{Internal threats.} First, our authorship decoupling heavily relies on the extraction of semantic relationships between Java/Kotlin packages. Tool and technical limitations can lead to the incorrect (or missing) relationships between pairs of classes. In this study, we use IC3-DialDroid to build ICC connections, and \textsc{AndroGuard} to identify call relations which may miss some implicit connections due to, for example, implicit Intent, reflection, and callbacks. Besides, some relations can only be discovered during execution, e.g., late binding. 
Second, not all components in an app are generated by the same author, \eg, repackage. This can affect module decoupling and the determination of the primary module. 
All experiments in this work were performed under the assumption that all code in the package was generated by the same author and that the primary module is where \textsf{\small MainActivity} resides. Android architect may also initialize its UI architecture and other developers develop its content part. The Android architecture has a little bit to do with the author's writing habits. 
Last, the elimination of legacy code from the Android framework is mainly based on a pre-defined list, which may be influenced by different Android SDKs or platforms. 
In future research, we intend to further refine our authorship decoupling approach, and develop a more robust approach to identify and track the legacy code of the Android framework.

\noindent\textbf{External threats.} External threats are more from the test datasets. 
First, Our test apps have a certain number of cloned apps, which may be caused by version upgrade. Although this cloned code is also attributed to the same author, it causes that any code clone detection technique can perform well. 
We have measured the influence of author number to authorship identification, as we think if there are more authors, the similarity of their code probably increases, raising the difficulty of authorship identification. According to the result in Section~\ref{sec:eval:authorship}, only 5\% drop occurs along with the increase of author number. However, it may be not able to represent the massive number of Android developers in the wild. Second, class imbalance may influence our result. For example, in dataset ANVA, there is one author that creates 103 apps, while another author has 17 apps. However, our approach proves to be performing very well in the other datasets.
Considering the above, we plan to collect more comprehensive apps and further compare them with relevant tools such as code clone detection and authorship identification. 
\section{Related Work}\label{sec:related}
\subsection{Android Static Analysis}

\noindent\textbf{Libraries Analysis.} Li \etal~\cite{libd} developed a prototypical tool called LibD, which utilizes code dependencies of Android apps to detect candidate libraries and handles multi-package third-party libraries in the presence of name-based obfuscation. Derr \etal~\cite{libscout} devised a light-weight and effective approach to detect third-party libraries that is resilient to common obfuscation techniques and capable of pinpointing exact library versions. Zhang \etal~~\cite{libid} presented a novel library detection tool named LibID, which is more resilient to code shrinking and package modification than state-of-the-art tools. 

\noindent\textbf{ICCs analysis.} Octeau \etal~\cite{ic3} designed an Android ICC analysis tool named IC3 to extract information related to the ICC sources, sinks and intent-based communication channels. Li \etal~\cite{iccta} proposed a taint analyzer named IccTA to detect privacy leaks among components in Android applications. Bosu \etal~\cite{DIALDroid} developed DIALDroid, an Android security tool for analyzing data flows between sensitive applications based on ICC. It leverages the relational database for a scalable matching of ICC entry and exit points, and fast analysis.

\subsection{Code Authorship Attribution}
Oman and Cook \cite{Oman} conducted a statistical analysis of authors' programming style and extracted typographic characteristics such as indentation, spacing, comments, and upper/lower cases of letters. Based on these characteristics, a clustering method is performed to successfully group the source code for several algorithms in textbooks as per authors. 
Frantzeskou \etal~\cite{Frantzeskou} proposed a SCAP (Source Code Author Profiles) approach based on byte-level characteristics. In particular, they extracted n-grams from source code including all non-printable characters, and used the highest n-grams as the marker for authors. 
Kalgutkar \etal~\cite{Kalgutkarauthorship} identified from Android apks all present strings including strings referenced by identifiers, string components, and strings in XML files. Then they built n-grams and leveraged SVM (Support Vector Machine) to classify the authors of code.
Caliskan \etal~\cite{Caliskan} identified the stylistic features from source code, \ie, lexical features, layout features and syntactic features. With random forest, they outperform the accuracy of authorship attributions than prior works. 
Meng \etal \cite{MengMJ17} presented four types of features \ie, instruction, control flow, data flow and context features. With a proposed joint classification model, they managed to identify the author for a single basic block in binary.
Abuhamad \etal~\cite{AbuhamadAMN18} proposed a Deep Learning-based Code Authorship Identification
System (DL-CAIS) for code authorship attribution. DL-CAIS proceeds with TF-IDF representation using deep neural networks and an author classifier based on Random Forest. \emph{Our study contributes to the contemporary authorship identification twofold. First, we propose authorship decoupling to separate code by different authors from the third-party libraries, compilation, Android framework, etc. To the best of our knowledge, none of the works in the Android field have considered this problem. On the other hand, we empirically analyze the noise brought by the Android system including the inherited classes and methods, invoked APIs, and try to solicit stylometric features that better characterize one author rather than the Android system. These can, from both theory and practice, benefit the researchers and practitioners in authorship identification of Android apps.}
\section{Conclusion}\label{sec:concl}

In this paper, we propose the \tool approach, which consists of two steps for authorship attribution. First, we conduct authorship decoupling through constructing package relation graph and cluster packages into different modules. In such a manner, the primary module can be further determined by the location where the entry point activity resides. In the course of authorship identification, we distill three types of features, retaining the stylometric features of app authors while removing imprints by the Android framework. 
An embedding algorithm and three types of machine learning algorithms are conducted to identify the leading authors of given apps. Our approach has been evaluated in four datasets, and the result shows that \tool can effectively identify authorship with an average accuracy of 92.8\% with Linear SVM, 92.4\% with Random Forest, and 92.4\% with Logistic Regression. It also proves that \tool performs still effectively on obfuscated code. 
Compared to an open-source authorship attribution tool, our approach makes a 3.4\% improvement in accuracy for authorship identification.
\section*{ACKNOWLEDGMENT}
We would like to thank the anonymous reviewers and shepherd for their valuable comments for this paper.
This work has partially been sponsored by the National Natural Science Foundation of China (No. 61872262, 61702045, 61572349, 61902395, and U1836211) and National Key R\&D Programmes of China (No. 2019AAA0104301). 
Xiaohong Li and Weimin Ge are the corresponding authors. 
\newpage
\balance
\bibliographystyle{IEEEtran}
\bibliography{main}
\end{document}